\documentclass{article}

\usepackage[margin=1.5cm, includefoot, footskip=30pt]{geometry}

\usepackage[T1]{fontenc}
\usepackage{amsmath}
\usepackage{booktabs}
\usepackage{graphicx}
\usepackage{hyperref}
\usepackage{listings}
\usepackage{lmodern}
\usepackage{multicol}
\usepackage{standalone}
\usepackage{xcolor}
\usepackage{subcaption}
\usepackage{makecell}
\usepackage{authblk}
\usepackage[symbol]{footmisc}
\usepackage[sort&compress,comma,numbers]{natbib}

\usepackage{tikz}
\usetikzlibrary{calc, shapes, patterns}

\usepackage{setspace}

\newcommand{\TFT}{\emph{TFT}}
\newcommand{\IPD}{\emph{IPD}}
\newcommand{\AXL}{\texttt{Axelrod-Python}}
\newcommand{\AXLFortran}{\texttt{Axelrod\_fortran}}
\newcommand{\SI}{\textbf{Supplementary Information}}

\title{Reviving, reproducing, and revisiting Axelrod's second tournament}
\author[1, $*$]{Vincent Knight}
\author[2]{Owen Campbell}
\author[3]{Marc Harper}
\author[4]{T. J. Gaffney}
\author[5, 6]{Nikoleta E. Glynatsi}

\affil[1]{School of Mathematics, Cardiff University}
\affil[2]{Independent Researcher, Cardiff, United Kingdom}
\affil[3]{Google Inc.}
\affil[4]{Independent Researcher, Nevada, USA}
\affil[5]{Discrete Event Simulations Teams, RIKEN Center for Computational Science}
\affil[6]{Mathematical Social Science Team, RIKEN Center for Interdisciplinary Theoretical and Mathematical Sciences}
\affil[$*$]{Corresponding author: KnightVA@cardiff.ac.uk.}
\date{}

\begin{document}

\maketitle

\begin{abstract}

Direct reciprocity, typically studied using the Iterated Prisoner's Dilemma
(\IPD), is central to understanding how cooperation evolves. In the 1980s, Robert
Axelrod organized two influential \IPD{} computer tournaments, where Tit for Tat
(\TFT) emerged as the winner. Yet the archival record is incomplete: for the
first tournament only a report survives, and for the second the submitted
Fortran strategies remain but not the final tournament code. This gap raises
questions about the reproducibility of these historically influential results.
We recreate the second tournament by restoring the surviving Fortran
implementations to compile with modern compilers and by building a Python
interface that calls the original strategy functions without modification. Using
the open-source \AXL{} library to run tournaments, we reproduce Axelrod's main
findings: \TFT{} prevails, and successful play tends to be cooperative, responsive
to defection, and willing to forgive. Strategy rankings remain mostly unchanged.
We then assess the robustness of the originally submitted strategies by
incorporating additional strategies, and we run one of the largest \IPD{} tournaments 
to date. We find that the original tournament was especially
favorable to \TFT{} and that it is difficult to dethrone \TFT{} when the original
submissions make up the majority of the field. We also observe that several
lesser-known submissions perform strongly in more diverse settings and
under noise.
Our contributions are: (i) the first systematic reproduction of Axelrod's second
tournament; (ii) a contemporary reassessment of the original results in light of
new strategies and settings; and (iii) a preserved, easy-to-use implementation
of the second-tournament strategies within \AXL{} to support future research.

\end{abstract}

\section{Introduction}\label{sec:introduction}

Cooperation is fundamental to the organization of societies, yet its
persistence remains a central puzzle in evolutionary biology and the social
sciences. In the absence of supporting mechanisms, natural selection favors
defection over cooperation. To explain the emergence of cooperative behavior,
several mechanisms have been proposed, among which \emph{direct reciprocity}
has played a central
role~\cite{trivers:QRB:1971,Nowak:Science:2006,Garcia:Frontiers:2018,Glynatsi:HSSC:2021,Rossetti:Ethology:2024}.
Direct reciprocity relies on repeated interactions among the same individuals,
where decisions can be conditioned on a partner's previous behavior. By
rewarding cooperation and punishing defection, such interactions provide a
pathway for cooperation to be maintained.

The standard model for direct reciprocity is the \emph{Iterated Prisoner's
Dilemma (IPD)}. In each round, two players independently decide whether to
cooperate (C) or defect (D). Mutual cooperation yields the reward payoff ($R$),
mutual defection gives the punishment payoff ($P$), and unilateral defection
results in the temptation payoff ($T$) for the defector and the sucker's payoff
($S$) for the cooperator. With $T>R>P>S$ and $2R>T+S$, the dilemma
arises: defection dominates in the one-shot game, even though mutual
cooperation would maximize collective welfare.

Because repeated interactions allow for a wide variety of conditional behaviors,
direct reciprocity gives rise to a diverse set of strategies. Much of
the field has therefore focused on understanding which strategies are
``successful'' and under what conditions. Researchers have examined which
strategies can constitute Nash equilibria~\cite{hilbe2018partners,
murase2020five, chen2023outlearning, glynatsi2024conditional}, which can emerge
and persist under evolutionary dynamics~\cite{van2012direct, baek2016comparing,
amaral2016stochastic, hilbe2017memory, Garcia:Frontiers:2018,
laporte2023adaptive}, and which perform best in computational
tournaments~\cite{li2007design, Stewart2012, carvalho2013iterated,
gaudesi2015exploiting, Knight2017, Glynatsi:PNAS:2024}.
Studies have also sought to identify the principles underlying
successful strategies, such as the balance between retaliation and forgiveness,
the role of stochasticity~\cite{baek2016comparing}, or the capacity to adapt to diverse opponents~\cite{glynatsi2024properties}.

A pioneering contribution to the study of direct reciprocity and strategies came from Robert
Axelrod's computer tournaments in the 1980s~\cite{Axelrod1980a,Axelrod1980b,Axelrodbook}.
Axelrod invited both academics and the wider public, including readers of
computer hobbyist magazines, to submit computer programs for playing the \IPD. Each
submission entered a \emph{round-robin tournament}, facing every other entry, a
copy of itself, and a random baseline strategy that cooperated with probability
$1/2$. Average payoffs across all matches determined the final ranking of
strategies. The first tournament~\cite{Axelrod1980a} attracted 14 entries and
was won by \emph{Tit for Tat (TFT)}, a remarkably simple strategy that mirrored
the opponent's previous move. A second, larger tournament~\cite{Axelrod1980b}
followed, with 63 strategies this time. The main change of the second tournament
was that matches no longer had a
fixed length but instead had 5 different lengths unknown to the players, removing the
possibility of exploiting a known horizon. Despite the new list of opponents, \TFT{}
again emerged as the winner. Axelrod argued that its success rested on five key
properties: it was \emph{nice} (never the first to defect), \emph{provocable}
(ready to retaliate), \emph{forgiving} (able to return to cooperation),
\emph{non-envious} (not striving to outperform the opponent), and \emph{simple}
(easy to understand and implement).

These pioneering tournaments shaped decades of research on cooperation, yet the
archival record surrounding them is incomplete. For the first tournament, only
the published report survives; the original source code has been lost. For the
second tournament, the Fortran implementation of the submitted strategies
remains available online,\footnote{See
\url{http://www-personal.umich.edu/~axe/research/Software/CC/CC2/TourExec1.1.f.html}
accessed on 2025-09-30.}
but the final code used in Axelrod's reported tournaments has not been preserved.

This highlights a broader challenge of software reproducibility in
the \IPD{} literature. Over the past 50 years, hundreds of strategies have been
proposed, often defined only by fragments of source code, incomplete
descriptions, or insufficient supporting data. This has made it difficult to
replicate earlier findings, and has led many researchers to evaluate new
strategies against small, unrepresentative sets of opponents. Such practices
bias conclusions and weaken claims about the relative performance of new
strategies.
An important step toward addressing this issue has been the
\texttt{Axelrod-Python} project~\cite{AxelrodProject}, an open-source Python
package that provides a comprehensive framework for implementing and testing
\IPD{} strategies. The library includes a wide variety of strategies from the
literature, together with detailed documentation and usage examples. By
providing open, executable implementations, \AXL{} makes it possible to test
strategies under common conditions and compare their performance systematically,
and it has therefore been used in ongoing research~\cite{Harper2017,
Knight2017, glynatsi2024properties, bucciarelli2024studying, macmillan2024game}.

One fundamental question has remained: \emph{can Axelrod's original results be
reproduced?} Addressing this question is important not only for assessing the
robustness of foundational research on cooperation, but also as a case study in
the reproducibility of computational social science more broadly. In this work, we
build on the \AXL{} project to replicate, as faithfully as possible, the
outcomes of Axelrod's \emph{second tournament}. Rather than manually
re-implementing the strategies from partial descriptions, which would risk
introducing errors, we update the surviving Fortran implementations so that they
compile with modern compilers, construct an interface that allows them to
interact with Python, and integrate them directly with the \AXL{} library. This
design allows us to call the original functions without modification.

Although Axelrod's reported results are impossible to replicate, we are able
to recover the main conclusions of Axelrod's study. At the same time, we extend
the analysis by incorporating additional strategies from the \AXL{} library and
ultimately conducting one of the largest \IPD{} tournaments to date. These
extensions shed new light on the robustness of the original findings. Overall,
this work makes three contributions. First, it provides the first systematic
attempt to reproduce the results of Axelrod's second tournament, which have been
central to the study of cooperation. Second, it offers a contemporary
perspective on Axelrod's experiments. Finally, by reviving Axelrod's
original code and making it accessible through the \AXL{} library, we preserve
for future use the strategies of his second tournament.

\section{Reviving the tournament}\label{sec:reviving}

As noted above, Axelrod's original tournaments cannot be fully replicated. For
the first tournament, the only surviving record is the published report, while
the source code has been lost. For the second tournament, however,
implementations of the 63 submitted strategies remain available. In this work,
we aim to revive the second tournament using these historical
implementations, rather than re-implementing them ourselves. The surviving code
was written in Fortran and is hosted online by the University of Michigan Center
for the Study of Complex Systems~\cite{Axelrod1980bCode}, where it was last
updated in 1996. The source code for all strategies is contained in a single file,
\texttt{TourExec1.1.f}, embedded in an HTML page.

Our first step was to ensure that this code could be executed with modern
software. We stripped the HTML formatting to recover the raw Fortran source and
applied only minimal modifications to enable compilation with contemporary
Fortran compilers. Each strategy was then extracted into its own modular file.
To streamline the build process, we created a dedicated \texttt{Makefile} that
compiles all strategy files into a single shared library
(\texttt{libstrategies.so}). Once installed in a standard location on a
POSIX-compliant operating system (e.g., \texttt{/usr/local/lib/}), this library
is automatically located at runtime. The modified source code is openly
available at \url{https://github.com/Axelrod-Python/TourExec} and has been
archived at~\cite{TourExec}, ensuring long-term accessibility.

Each Fortran strategy was implemented as a function with a fixed set of input
arguments that encode the state of the match:

\begin{itemize}
    \item \texttt{J}: Opponent's previous move (0 = cooperate, 1 = defect),
    \item \texttt{M}: Current turn number (starting at 1),
    \item \texttt{K}: Player's cumulative score,
    \item \texttt{L}: Opponent's cumulative score,
    \item \texttt{R}: Random number between 0 and 1 (for stochastic strategies),
    \item \texttt{JA}: Player's previous move.
\end{itemize}

Given these inputs, the function outputs either 0 (cooperate) or 1 (defect). An
illustrative example is provided by the original implementation of \TFT{}
(\texttt{k92r.f}), shown in Figure~\ref{fig:tft_fortran}.

\begin{figure}[!hbtp]
    \begin{center}
        \begin{lstlisting}[language=Fortran,
                           basicstyle=\ttfamily\scriptsize,
                           frame=single,
                           keywordstyle=\color{red}\bfseries,
                           commentstyle=\color{teal}\itshape]
          FUNCTION K92R(J,M,K,L,R,JA)
    C BY ANATOL RAPOPORT
    C TYPED BY AX 3/27/79 (SAME AS ROUND ONE TIT FOR TAT)
    c replaced by actual code, Ax 7/27/93
    c  T=0
    c   K92R=ITFTR(J,M,K,L,T,R)
          k92r=0
          k92r=j
    c test 7/30
    c   write(6,77) j, k92r
    c77   format(' test k92r. j,k92r: ', 2i3)
          RETURN
          END
        \end{lstlisting}
    \caption{\textbf{Fortran source code for the Tit For Tat strategy
    (\texttt{k92r.f}).} Each strategy takes as input the state of the match
    (\texttt{J}, \texttt{M}, \texttt{K}, \texttt{L}, \texttt{R}, \texttt{JA})
    and outputs either 0 (cooperate) or 1 (defect). This code corresponds to
    the \TFT{} strategy submitted to Axelrod's second tournament.}
    \label{fig:tft_fortran}
    \end{center}
\end{figure}

With the strategies executable again, the next challenge was to run the
tournament. For this we rely on the open-source \AXL{} library, which provides a
modern implementation of \IPD{} tournaments. In this framework, a
\emph{tournament} is a collection of matches, and each \emph{match} is a
repeated game between two strategies. A tournament class manages the overall
structure (pairing strategies, scheduling matches, aggregating results), while
strategy classes determine actions turn by turn based on the recorded history of
play.

\begin{figure}[!hbtp]
    \begin{center}
        \includegraphics[width=.7\textwidth]{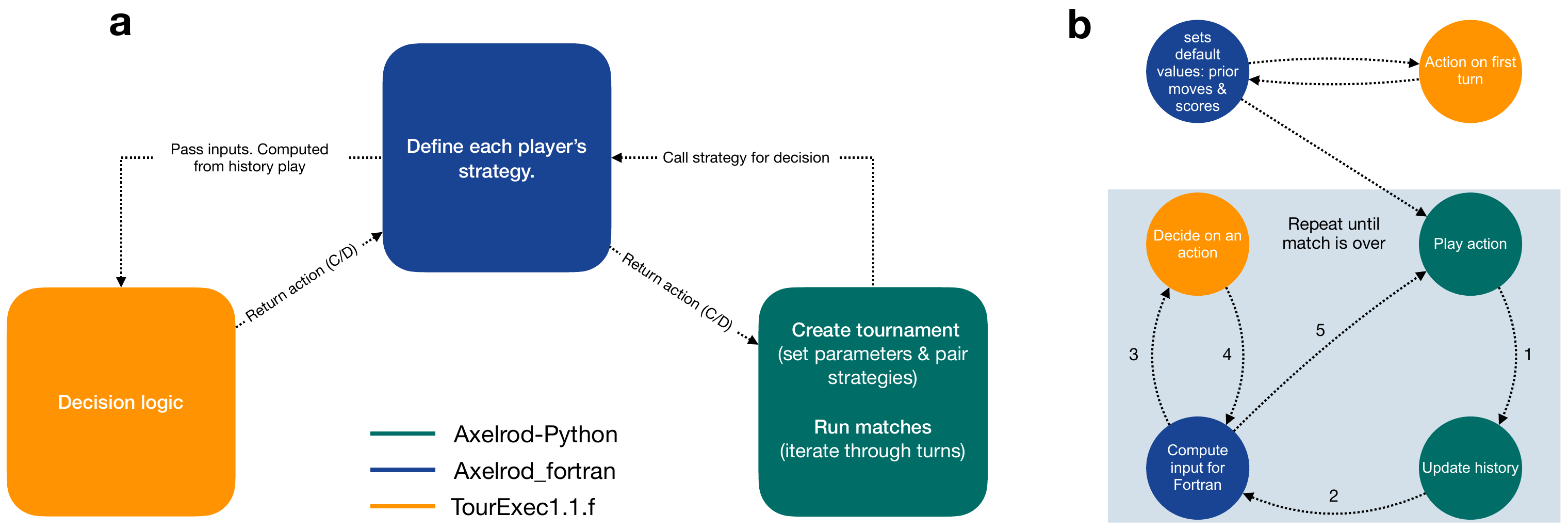}
\caption{\textbf{Diagrammatic representation of the interaction between \AXL{},
\AXLFortran{}, and the original Fortran code.} \textbf{a,} Overview of how the
three codebases interact. \AXLFortran{} creates strategy instances for each
player. Tournaments and matches are managed by \AXL{}, which at each turn calls
the corresponding \AXLFortran{} instance; this in turn invokes the original
Fortran code to produce a decision from the match history. \textbf{b,} Turn-by-turn execution.
On the first turn, \AXLFortran{}
initialises the parameters required by the Fortran implementation, queries it
for the first decision, and returns this to \AXL{}, which plays the move and
updates the history. On each subsequent turn: (1) the previous turn is recorded
and the history updated; (2) \AXLFortran{} access the updated history and (3) 
formats its into the inputs required by the Fortran code; (4) the Fortran code is called and returns
a decision; (5) \AXL{} plays the decision. The process repeats until the match
ends.}\label{fig:strategy_diagram}
    \end{center}
\end{figure}

To integrate the original Fortran strategies into this environment, we developed
a Python package, \AXLFortran~\cite{Axelrod_fortran}. The \AXLFortran{} adapter
class inherits from the base \AXL{} strategy class and translates between the two
languages: it constructs the required inputs, calls the Fortran function, and
returns the chosen move to the Python environment
(Figure~\ref{fig:strategy_diagram}). The adapter also handles the initial moves
of both players
required by the Fortran strategies, which is fixed to cooperation in the original
code. 

The major advantage of this approach is that no subjective reinterpretation was
introduced in reproducing Axelrod's second tournament. The only modifications
made were minimal adjustments to ensure compatibility with modern compilers. For
several strategies, the Fortran source code is the only surviving description;
these strategies are therefore run and evaluated exactly as originally written.
This constitutes the most faithful reproduction of Axelrod's
work to date.

\section{Reproducing Axelrod's second tournament}\label{sec:reproducing}

\textbf{Running the tournament.}
Axelrod's second tournament included 63 strategies. While Axelrod
named a few explicitly, most were referred to only by their ranking in the
original tournament or by the names of the individuals who submitted them. We
now have access to the function names of the Fortran implementations. In what
follows, we refer to strategies by their function names, supplemented by
Axelrod's names when available. A complete list of the strategies, their
authors, their original rankings, the names provided by Axelrod (where
available), and their function names is given in the \SI~(S1).

Axelrod use the same payoff matrix as in the first tournament, with
\(T=5\), \(R=3\), \(P=1\), and \(S=0\).
Unlike the first tournament, players were not informed of the number of rounds
in advance. In~\cite{Axelrod1980b} it is stated that:

\begin{quote}
    \textit{``As announced in the rules, the length of the games was determined
        probabilistically with a .00346 chance of ending with each given move.
        This parameter was chosen so that the expected median length of a game
        would be 200 moves. In practice, each pair of players was matched five
        times, and the lengths of these five games were determined once and for
        all by drawing a random sample. The resulting random sample from the
        implied distribution specified that the five games for each pair of
        players would be of lengths 63, 77, 151, and 308 moves. Thus the average
        length of a game turned out to be somewhat shorter than expected at 151
        moves.''}
\end{quote}

Thus, termination was not determined probabilistically during play. Instead,
Axelrod sampled five match lengths in advance of the tournament and applied the
same set to every pair of players. The text lists only four lengths
(\(63, 77, 151, 308\)) even though five games were played. Using the reported
average length of 151 moves, the missing length can be inferred to be \(156\),
so we assume the complete set of match lengths to be \(\{63, 77, 151, 156, 308\}\).

It also appears that the only mechanism for averaging out stochastic variation in
the original setup was the use of these five fixed repetitions. Without access to
the random seed used by Axelrod, it is not possible to reproduce the tournament
exactly. To obtain smoother estimates of each strategy's performance, we run
Axelrod's second tournament a total of 25,000 times.
For each run of the tournament, we use the same five match lengths
(\(63, 77, 151, 156, 308\)) for every pairing of strategies.

\begin{figure}[!hbtp]
    \centering
    \includegraphics[width=.85\textwidth]{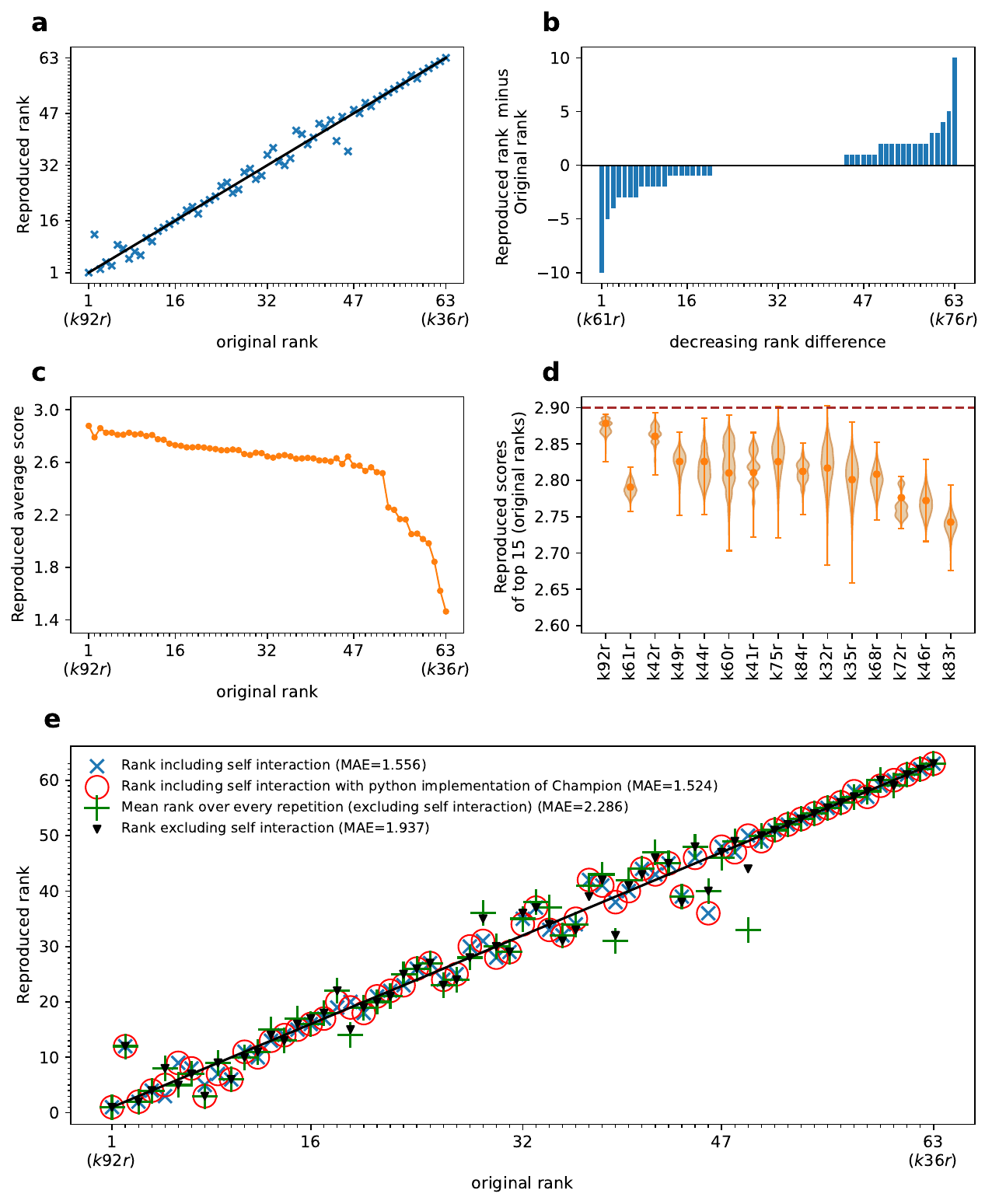}
\caption{\textbf{Reproducing Axelrod's second tournament.}
\textbf{a,} Comparison of the original rankings and the replicated rankings.
Strategies are ordered by their original ranks; \TFT{} again
emerges as the overall winner, while \texttt{k36r} again ranks last.
\textbf{b,} Differences in rank between the original and reproduced tournaments,
ordered from most negative to most positive change. Most strategies shift by
only 1-3 positions, though a few exhibit larger changes, including Champion
(\texttt{k61r}).
\textbf{c,} Mean scores of all strategies. Champion's
average score drops relative to the other three strategies but remains close to
the rest of the top ten.
\textbf{d,} Violin plots showing the distribution of scores for the top 15
strategies across all repetitions.
\textbf{e,} Results under alternative replication approaches, including with and
without self-interactions and using a Python re-implementation of Champion.
Across all settings, \TFT{} remains the winner, while Champion exhibits the
largest change in ranking.}
  \label{fig:replicated_tournament}
\end{figure}

\noindent
\textbf{The winners, the losers, and the differences.}
In Figure~\ref{fig:replicated_tournament}\textbf{a}, we compare the original rankings with
those obtained in our replication, with strategies sorted according to their
original ranks. As we can see, the winner of the tournament remains \TFT, while the strategy
\texttt{k36r}, authored by Roger Hotz, again ranks last.
The overall outcomes of
most strategies are reproduced, though some do not retain their exact rankings.
Notably, the bottom ranking strategies are replicated quite well, consistently
performing poorly in both tournaments. Figure~\ref{fig:replicated_tournament}\textbf{b}
shows the difference in rank between the reproduced and original tournaments,
ordered from the most negative to the most positive change. We find that 40 out
of 63 strategies exhibit a shift in ranking. Most of these changes are modest,
typically within 1-3 positions, though two strategies move by 10 places.

These two strategies are \texttt{k76r} (originally ranked 46th, now 36th) and
\texttt{k61r} (originally 2nd, now 12th). The latter, \texttt{k61r}, is also
known as \emph{Champion}, after its author's surname. Champion cooperates for the
first ten moves, plays \TFT{} for the next fifteen, and thereafter cooperates
unless (i) the opponent defected on the previous move, (ii) the opponent's
overall cooperation rate falls below 60\%, and (iii) a random draw in $[0,1]$
exceeds the opponent's cooperation rate.
Upon closer inspection, we identified a bug in the Fortran implementation of
Champion: an internal variable was not properly initialized. Specifically, the
variable \texttt{ICOOP} was not assigned an initial value within the function. On the very
first call \texttt{ICOOP} is effectively treated as $0$, but in later matches this
assumption no longer holds, meaning that while in the first match \texttt{ICOOP}
starts at $0$ and increases as expected, in subsequent matches its (undefined)
value can persist across calls. The corrected version of this strategy is shown in
Figure~\ref{fig:k61r}. It is unclear whether this bug affected the tournament
results reported in 1980. One possibility is that each match was run in isolation,
which would have prevented the error from arising. In our analysis we use the
corrected version of the strategy.

In Figure~\ref{fig:replicated_tournament}\textbf{c}, we examine the average scores of the
strategies. Champion shows a noticeable drop relative to the now
second and third ranked strategies. However, its mean score remains very close to that
of the rest of the top ten (Table~\ref{tbl:original_tournament_rankings}).
Figure~\ref{fig:replicated_tournament}\textbf{d} presents violin plots of the
score distributions for these top 15 strategies across the
25,000 repetitions. Champion
rarely outperforms the second or third ranked strategies, doing so in only a
single tournament, confirming that, at least in this version, Champion
is not as strong as originally reported.

\begin{table}[!hbtp]
        \centering
        \resizebox{.9\textwidth}{!}{
        \begin{tabular}{llcccc}
\toprule
 & \makecell[c]{Author} & \makecell[c]{Average Score} & \makecell[c]{Reproduced Rank} & \makecell[c]{Original Rank} & \makecell[c]{Average\\Cooperation Rate} \\
\midrule
k92r & Anatol Rapoport & 2.878 & 1 & 1 & 0.922 \\
k61r & Danny C Champion & 2.791 & 12 & 2 & 0.954 \\
k42r & Otto Borufsen & 2.861 & 2 & 3 & 0.916 \\
k49r & Rob Cave & 2.826 & 4 & 4 & 0.892 \\
k44r & William Adams & 2.826 & 3 & 5 & 0.882 \\
k60r & Jim Graaskamp and Ken Katzen & 2.810 & 9 & 6 & 0.844 \\
k41r & Herb Weiner & 2.811 & 8 & 7 & 0.865 \\
k75r & Paul D Harrington & 2.826 & 5 & 8 & 0.802 \\
k84r & T Nicolaus Tideman and Paula Chieruzz & 2.812 & 7 & 9 & 0.882 \\
k32r & Charles Kluepfel & 2.817 & 6 & 10 & 0.873 \\
k35r & Abraham Getzler & 2.801 & 11 & 11 & 0.888 \\
k68r & Fransois Leyvraz & 2.808 & 10 & 12 & 0.917 \\
k72r & Edward C White Jr & 2.776 & 13 & 13 & 0.924 \\
k46r & Graham J Eatherley & 2.772 & 14 & 14 & 0.948 \\
k83r & Paul E Black & 2.743 & 15 & 15 & 0.935 \\
\bottomrule
\end{tabular}
}
        \caption{\textbf{Top 15 strategies in the reproduced tournament.}
        The table lists the top 15 strategies from Axelrod's original tournament
        along with their rankings in our tournament. For each strategy we also
        report the average score and average cooperation rate in the reproduced
        tournament.}
        \label{tbl:original_tournament_rankings}
\end{table}

An interesting observation is that strategies such as \texttt{k32r} and
\texttt{k75r} occasionally achieve the maximum possible score of 2.9. Across all
repetitions, these strategies won 0.8\% and 2.3\% of the tournaments,
respectively (Table~\ref{tbl:original_tournament_win_proportion}).
They received little attention in the original tournament, likely because their
high variance obscured their potential relative to consistently strong
performers such as \TFT{}. Another strong and comparatively stable strategy is
\texttt{k42r}, which won 31.5\% of our replicated tournaments; for comparison,
\TFT{} won 63.3\% (Table~\ref{tbl:original_tournament_win_proportion}).
Because the original tournament is a single realization, a different random seed
could plausibly have produced a different winner: namely, on another roll of the
dice, \texttt{k42r} (Otto Borufsen) might well have been crowned the winner.

\begin{table}[!hbtp]
        \centering
        \begin{tabular}{lr}
\toprule
 & Win proportion \\
\midrule
k32r & 0.0079 \\
k35r & 0.0002 \\
k42r & 0.3136 \\
k44r & 0.0038 \\
k60r & 0.0103 \\
k75r & 0.0226 \\
k78r & 0.0016 \\
k82r & 0.0083 \\
k92r & 0.6319 \\
\bottomrule
\end{tabular}

        \caption{\textbf{The proportion of reproduced tournaments won by each
        strategy.}
        The table shows all strategies who won one of the 
        repetitions of the
        original tournament. \TFT{} wins 63.3\% of them indicating that 36.7\% of
        used random seeds lead to a different winner.}
        \label{tbl:original_tournament_win_proportion}
\end{table}

The source code for \texttt{k42r}, together with a description of the strategy's
behaviour, is provided in the \SI~(S6). In summary, the strategy has two states:
\emph{normal} and \emph{defecting}. In its normal state, it behaves much like
\TFT{}, but it includes mechanisms to recover from mutual defection and to
escape alternating defections. Every 25 turns, it evaluates the opponent's
cooperation rate and switches to defecting mode if the opponent appears random
or highly defective. The \texttt{k42r} strategy can thus be viewed as a more
sophisticated and adaptive variant of \TFT{}. These mechanisms are likely
intended to prevent endless retaliation and to guard against exploitation by
less cooperative opponents. However, in Axelrod's original tournament,
\texttt{k42r} ranked only 3th, suggesting that these mechanisms were not
particularly advantageous/or used against the original pool of strategies
and highly cooperative environment.

We also explored alternative approaches in our attempt to replicate the
original ranking (Figure~\ref{fig:replicated_tournament}\textbf{e}). These include running
the tournament with and without self-interactions, as well as testing a Python
implementation of Champion that we developed from the textual
description in~\cite{Axelrod1980b}. Across all approaches, Champion
consistently exhibits the largest change in ranking among the top strategies,
while \TFT{} remains the overall winner.

Overall, the minor discrepancies in rankings and scores can be attributed to one
or more of the following:

\begin{itemize}
\item An unknown modification of the source code of Champion (\texttt{k61r})
used in the final original tournament code~\cite{Axelrod1980b}.
\item Stochastic variation not being sufficiently taken in to account in the
original tournament~\cite{Axelrod1980b}.
\item A difference with how an older Fortran compiler would interpret the
commands, though the implemented version seems to
interact as expected.
\end{itemize}

\begin{figure}[!hbtp]
    \begin{center}
        \begin{lstlisting}[language=Fortran,
                           basicstyle=\ttfamily\scriptsize,
                           frame=single,
                           keywordstyle=\color{red}\bfseries,
                           commentstyle=\color{teal}\itshape]
      FUNCTION K61R(ISPICK,ITURN,K,L,R, JA)
C BY DANNY C. CHAMPION
C TYPED BY JM 3/27/79
      k61r=ja    ! Added 7/27/93 to report own old value
      IF (ITURN .EQ. 1) ICOOP = 0  ! Added 10/8/2017 to fix bug for multiple runs
      IF (ITURN .EQ. 1) K61R = 0
      IF (ISPICK .EQ. 0) ICOOP = ICOOP + 1
      IF (ITURN .LE. 10) RETURN
      K61R = ISPICK
      IF (ITURN .LE. 25) RETURN
      K61R = 0
      COPRAT = FLOAT(ICOOP) / FLOAT(ITURN)
      IF (ISPICK .EQ. 1 .AND. COPRAT .LT. .6 .AND. R .GT. COPRAT)
     +K61R = 1
      RETURN
      END
        \end{lstlisting}
    \caption{\textbf{Fortran source code for \texttt{k61r} (Champion).}
    The original implementation contained a bug: the variable \texttt{ICOOP} was not
    initialized within the function. In the first match it effectively started at 0,
    but in subsequent matches it could have kept its value from the previous match or be reset to 0,
    depending on the compiler and runtime environment. Line~5 shows the fix, which
    explicitly initializes \texttt{ICOOP} at the start of each match.}
        \label{fig:k61r}
    \end{center}
\end{figure}

\noindent
\textbf{The representative strategies.}
Axelrod conducted an analysis to identify what he termed ``representative
strategies''~\cite{Axelrod1980b}. These were strategies that, according to a
linear regression model, served as strong predictors of overall performance
(average score). The representative strategies identified in the original
tournament were S$_6$, Reversed State Transition, Ruler, Tester, and Tranquilizer.
Using the scores against these five strategies as predictors, Axelrod reported
an \(R^2\) value of 0.979,
indicating that 97\% of the variance in overall
performance could be explained by the model. For completeness, the coefficients
of this regression (as reported in~\cite{Axelrod1980b}) are listed in
Table~\ref{tbl:original_tournament_representative_model}.

We replicated Axelrod's linear regression analysis and evaluated the predictive
power of the original model using our reproduced tournament data. We
obtained an \(R^2\) of~0.957\unskip,
demonstrating that
the model continues to be a strong predictor of performance despite the small
discrepancies in individual rankings.

To further assess the robustness of the model, we performed a backward
elimination procedure to examine how \(R^2\) changes as the number of predictor
strategies increases. We find that the largest gain occurs when five predictors
are included, after which improvements are small. This suggests that
Axelrod's choice of five representative strategies was reasonable. Whether this
selection was the outcome of a systematic feature selection procedure or a
judgment call remains unclear; in either case, it appears well chosen. The
results of this elimination analysis are reported in the \SI~(S5).

\begin{table}[!hbtp]
\centering
\begin{tabular}{lc|cll}
\toprule
Strategies & \makecell{Originally Reported\\Coefficients} & \makecell{Replicated\\Coefficients} & $p$-value & $F$-value\\
\midrule
k60r (S$_6$) & 0.202000 & 0.232010 & 0.000000 & 298.163936 \\
k91r (Revered State Transition) & 0.198000 & 0.187930 & 0.000000 & 56.667196 \\
k40r (Ruler) & 0.110000 & 0.071870 & 0.000000 & 78.191831 \\
k76r (Tester) & 0.072000 & 0.063670 & 0.025992 & 5.207451 \\
k67r (Tranquilizer) & 0.086000 & 0.113650 & 0.000002 & 27.893148 \\
Intercept & 0.795000 & 0.790880 & NA & NA \\
\bottomrule
\end{tabular}
\caption{\textbf{The representatives and their effect on average score.} Linear
model described in~\cite{Axelrod1980b} with \(R^2=0.979\), Fitting a new model
to the same 5 strategies gives the coefficients gives
\(R^2=\protect 0.957\).}
\label{tbl:original_tournament_representative_model}
\end{table}

\begin{figure}[!hbtp]
    \centering
    \includegraphics[width=\textwidth]{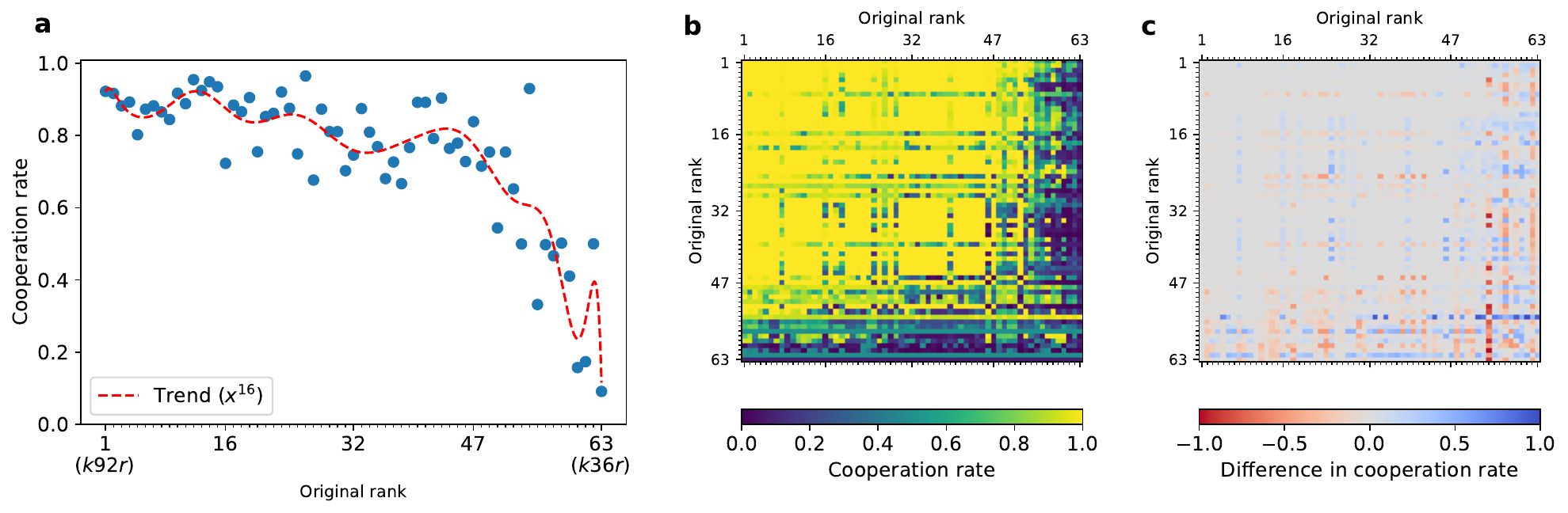}
\caption{\textbf{Cooperation in Axelrod's second tournament.}
\textbf{a,} Mean cooperation rate by original ranking. Most strategies are highly
cooperative, with the exception of those at the bottom of the ranking, which fall
below 50\%. \textbf{b,} Cooperation rates between each pair of players, players
are ordered by their original ranking. \textbf{c,} Deviation from symmetry in
pairwise cooperation rates, defined as the difference between cooperation matrices
and their transposes. Values close to zero indicate that both strategies cooperate
with one another at the same rate.}
  \label{fig:original_tournament_cooperation}
\end{figure}

\noindent
\textbf{Cooperation in the original tournament.}
Here we ask: \emph{what was the original tournament like?} The answer is that it
was, in fact, highly cooperative. The mean cooperation rate across the
tournaments is 75\%. When we plot cooperation rates against the original
rankings (Figure~\ref{fig:original_tournament_cooperation}\textbf{a}), most strategies are
indeed highly cooperative, with the exception of those in the bottom 14\% of the
ranking, which display cooperation rates below 0.5.

In Figure~\ref{fig:original_tournament_cooperation}\textbf{b}, we show that the
top performing strategies cooperate almost perfectly with one another, one of
Axelrod's key observations. He also emphasized the importance of
\emph{provocability} (willingness to punish defection) and \emph{forgiveness}
(ability to return to cooperation). In the case of \TFT, this means responding
immediately to the opponent's last move: \TFT{} retaliates after defection and
resumes cooperation after cooperation. As a result, \TFT{} achieves
the same cooperation rate as its opponent. To capture this notion, in
Figure~\ref{fig:original_tournament_cooperation}\textbf{c} we plot the difference between
pairwise cooperation rates and their transposes, that is, the deviation from
symmetry. A value of 0 indicates that two strategies cooperate at exactly the
same rate with one another. The top performing strategies cluster near 0,
whereas lower performing strategies are more asymmetric, with larger positive or
negative deviations.

These outcomes are not surprising. By the time of the second tournament,
Axelrod's earlier results were already known, and the strong performance of
\TFT{} was widely recognized. As a consequence, many newly submitted strategies
were explicitly designed to resemble \TFT{} or to cooperate effectively with
\TFT{}-like rules, hence their high levels of cooperation and the
symmetry in their interactions.

\section{Revisiting the tournament}\label{sec:extending}

\noindent
\textbf{Extra invitations.}
We investigate how the results would change if Axelrod had received additional
submissions. The \AXL{} library includes
209 strategies that did not appear in the original tournament. We therefore examine the
effect of including new strategies. We do this by first simulating a tournament
with all 
272 strategies
repeating each match-length 250 times. This reduced number of repetitions was
due to the combinatorial difficulty of running such a large tournament. This
full tournament will be discussed subsequently.
We use the interactions
from this full tournament to see the effect of
first adding a single new strategy, then considering all possible pairs, triples,
and quadruples drawn from the remaining pool. This yields a total of
78{,}760{,}605 tournaments:

\begin{enumerate}
    \item Adding 1 strategy: $\binom{209}{1}=209$ tournaments.
    \item Adding 2 strategies: $\binom{209}{2}=21{,}736$ tournaments.
    \item Adding 3 strategies: $\binom{209}{3}=1{,}499{,}748$ tournaments.
    \item Adding 4 strategies: $\binom{209}{4}=77{,}238{,}876$ tournaments.
\end{enumerate}

Our analysis focuses on how the additional submissions affect the performance of
the \emph{original} 63 strategies. For the original strategies that win at least
one of these new tournaments,
we show the probability of
winning the tournament when 1, 2, 3, or 4 additional strategies are included
(Table~\ref{tbl:alternate_extra_strategy_tournament_winner_proportion}). These
include \TFT{}, and high-scoring strategies such as \texttt{k75r},
\texttt{k32r}, and \texttt{k42r}.

A striking pattern is that the effect of extra invitations on \TFT{} is
\emph{non-monotonic}. Adding a single additional invitee substantially
\emph{increases} the probability that \TFT{} wins (to about 85\%); with two
invitees the probability is lower (\(\sim\)71\%), with four invitees it is
higher again (\(\sim\)75\%), and only with three invitees do we observe a drop
(to \(\sim\)60\%). Overall, these results highlight the robustness of \TFT{} in
Axelrod's second tournament: it is difficult to unseat as the top performer.
The next most successful strategy, \texttt{k42r}, remains competitive (e.g.,
winning up to \(\sim\)37\% with three additional invitees), whereas strategies
such as \texttt{k32r} and \texttt{k75r}, which can achieve high mean scores in
some realizations, tend to be even less effective as more entrants are introduced.

\begin{table}[!hbtp]
        \centering
        \resizebox{.9\textwidth}{!}{
        \begin{tabular}{lrrrr}
\toprule
 & 1 New (N = 209) & 2 New (N = 21736) & 3 New (N = 1499784) & 4 New (N = 77238876) \\
\midrule
k32r & 0.00000 & 0.00000 & 0.00001 & 0.00079 \\
k41r & 0.00000 & 0.00000 & 0.00001 & 0.00000 \\
k42r & 0.14833 & 0.26941 & 0.36640 & 0.21723 \\
k44r & 0.00000 & 0.00023 & 0.00057 & 0.00461 \\
k49r & 0.00000 & 0.00014 & 0.00035 & 0.00023 \\
k60r & 0.00478 & 0.01118 & 0.01882 & 0.00451 \\
k75r & 0.00000 & 0.00051 & 0.00245 & 0.00086 \\
k92r & 0.84689 & 0.71747 & 0.60814 & 0.75412 \\
Sum & 1.00000 & 0.99894 & 0.99674 & 0.98235 \\
\bottomrule
\end{tabular}
}
        \caption{\textbf{Proportion of wins for original strategies with 1-4 additional invitees.}
        Shown are the strategies with non zero probabilities of winning at least
        one of the
        modified tournaments under
        five fixed match lengths, repeated
        250 times. Additional
        invitees are strategies drawn from the open-source \AXL{} library; a
        full list is provided in the \SI{}, and source code/documentation are
        referenced therein. We consider four cases: adding 1, 2, 3, or 4 extra
        strategies to the original field. The last row reports the sum across
        these strategies (noting that some wins accrue to strategies not
        submitted to the original tournament).}
        \label{tbl:alternate_extra_strategy_tournament_winner_proportion}
\end{table}

\noindent
\textbf{A noisy tournament.}
A natural question is how the results would change if the original tournament
had included \emph{noise}, i.e., a small probability that a player's intended
action is flipped. Noise was among the main criticisms of Axelrod's study, since
\TFT{} is not robust to accidental defections~\cite{molander1985optimal,
bendor1993uncertainty, wu1995cope, do2017combination}.

\begin{figure}[!hbtp]
    \centering
    \includegraphics[width=.88\textwidth]{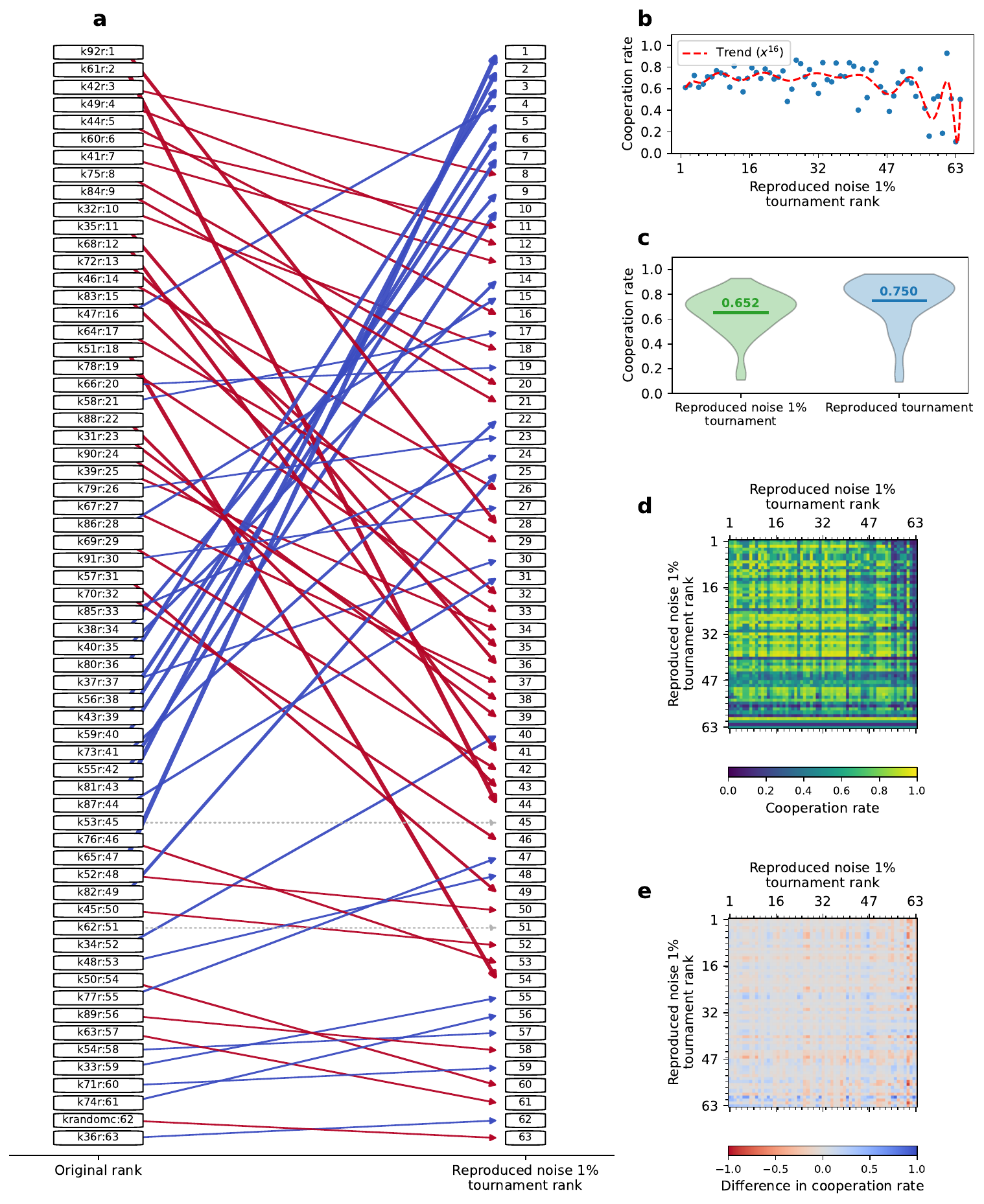}
    \caption{\textbf{Axelrod's second tournament with 1\% noise.}
    \textbf{a,} Comparison of the original rankings and the rankings of the second
    tournament with noise. In the left column, strategies are ordered by their
    original ranks, and arrows indicate their ranks in the noisy tournament (right
    column). The arrow widths are proportional to the change in rank. Blue arrows
    indicate a positive change in rank, red arrows indicate a negative change in
    rank, and grey arrows indicate no change in rank.
    \textbf{b,} Cooperation rates of strategies ordered by rank in this tournament.
    \textbf{c,} Tournament cooperation rates across all repetitions for
    this tournament (green) and the reproduced tournament without noise (blue).
    We also plot the average cooperation rate (green or blue line) and report its value.
    \textbf{d,} Pairwise cooperation rates between strategies. For each
    pair we calculate the cooperation rate of the row strategy towards the column
    strategy and vice versa. The strategies are ordered by the ranks in this
    tournament. For a full list of strategies see the \SI.
    \textbf{e,} Differences in cooperation rates for each pair of strategies, again 
    ordered by their rank in the reproduced tournament with noise.}
    \label{fig:reproduced_with_noise}
\end{figure}

Figure~\ref{fig:reproduced_with_noise} summarizes the results for a noise probability of
\(0.01\). Figure~\ref{fig:reproduced_with_noise}\textbf{a} shows rank changes among the 63 original
submissions: the left column orders strategies by their original rank, while
the right column lists ranks \(1\)-\(63\), with arrows indicating each strategy's
rank in the noisy tournament. As expected, noise leads to substantial reordering,
with several mid ranking strategies moving into top positions and some former top performers
dropping. The mean cooperation rate declines from \(0.75\) (noiseless) to \(0.65\)
under noise (Figure~\ref{fig:reproduced_with_noise}\textbf{c}). Correspondingly, the new
winning strategy in this setting cooperates at \(0.61\), slightly \emph{below} the
noisy-tournament mean, whereas \TFT{} cooperates at \(0.70\), \emph{above} the mean.
This pattern reinforces the idea that \emph{cooperating slightly less than the
tournament average} is a good predictor of success when noise is present~\cite{glynatsi2024properties}.
Figures~\ref{fig:reproduced_with_noise}\textbf{d} and \textbf{e} show that pairwise cooperation
becomes more asymmetric and generally lower under noise. Strategies that tend to
cooperate less than their co-players perform better. More complex strategies with
additional mechanisms also fare well; for example, \texttt{k42r}, now ranked 10th,
outperforms \TFT{}.

We emphasize that evaluating the original submissions under noise is not
entirely fair: many authors would likely have incorporated explicit
error-correction or forgiveness mechanisms had noise been part of the original
rules.

\noindent
\textbf{Large tournaments.}
In this section, we examine how the original strategies perform in larger and more
diverse tournaments. First, we combine the original submissions with the strategies
from the Stewart and Plotkin (S \& P) tournament~\cite{Stewart2012}. Next, we run
three large-scale tournaments using the full \AXL{} library under three noise
regimes: no noise, \(1\%\) noise, and \(5\%\) noise. We report the outcomes for
each tournament separately. Full results, including all rankings and cooperation
rates, are provided in the \SI.

\begin{figure}[!hbtp]
    \centering
    \includegraphics[width=.90\textwidth]{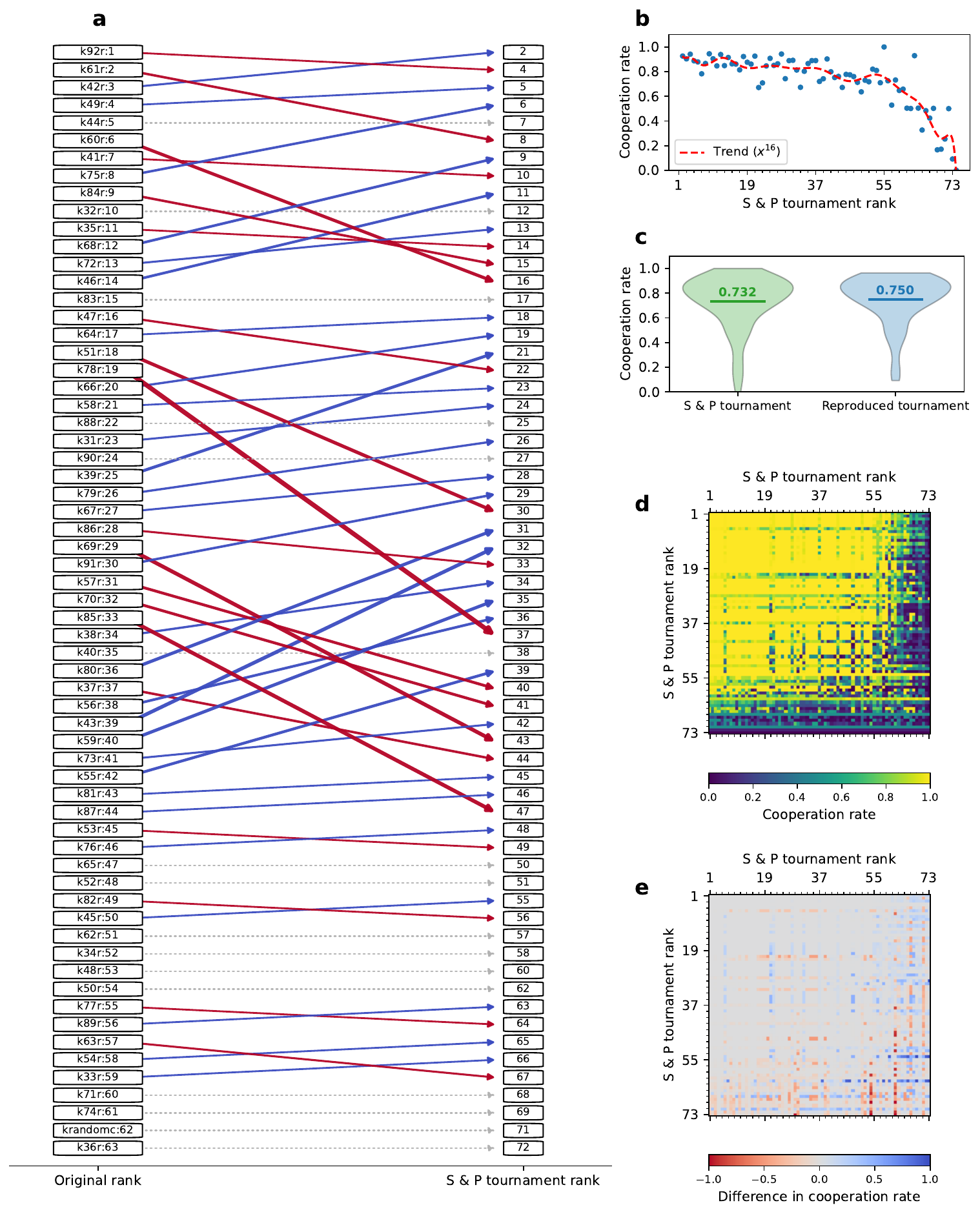}
    \caption{\textbf{Reproduction of Axelrod's second tournament with the addition
    of the strategies from Stewart's and Plotkin's tournament~\cite{Stewart2012}.}
    Similar to Figure~\ref{fig:reproduced_with_noise}. The right column shows the
    rankings on the original strategies in this new tournament.}
  \label{fig:rank_change_in_sp_tournament}
\end{figure}

Since the work of Press and Dyson \cite{Press2012}, there has been sustained
interest in zero-determinant (ZD) strategies. Stewart and Plotkin
\cite{Stewart2012} presented a focused tournament pitting ZD strategies against
a small collection of other rules. When all S \& P strategies are added to
Axelrod's second tournament, the resulting relative rank changes for the original
strategies are modest (Figure~\ref{fig:rank_change_in_sp_tournament}). The overall
level of cooperation remains high (mean \(\approx 0.73\)), and the leading
strategies are highly cooperative with one another, mirroring the qualitative
pattern of the original tournament. Table~\ref{tbl:sp_tournament_rankings_top_fifteen}
lists the top-performing strategies (see the \SI~(S3) for the full ranking). In
this new tournament, ZD strategies are distributed across the ranking rather than
concentrated at the top (\SI~(S3)), while \TFT{}, \texttt{k42r}, and generous variants
(e.g., GTFT~\cite{nowak1992tit} and ZD-GTFT-2) perform particularly well.

\begin{table}[!hbtp]
        \centering
        \resizebox{.8\textwidth}{!}{
        \begin{tabular}{lccccc}
\toprule
 & \makecell[c]{Average\\Score} & \makecell[c]{Rank} & \makecell[c]{Average\\Cooperation Rate} & \makecell[c]{Reproduced Average\\Cooperation Rate} & \makecell[c]{S. and P. Average\\Cooperation Rate} \\
Name &  &  &  &  &  \\
\midrule
ZD-GTFT-2: 0.25, 0.5 & 2.834 & 1 & 0.926 & - & 0.855 \\
k42r & 2.825 & 2 & 0.903 & 0.916 & - \\
GTFT: 0.33 & 2.819 & 3 & 0.941 & - & 0.893 \\
k92r & 2.812 & 4 & 0.888 & 0.922 & 0.731 \\
k49r & 2.787 & 5 & 0.879 & 0.892 & - \\
k75r & 2.786 & 6 & 0.782 & 0.802 & - \\
k44r & 2.781 & 7 & 0.866 & 0.882 & - \\
k61r & 2.773 & 8 & 0.945 & 0.954 & - \\
k68r & 2.768 & 9 & 0.904 & 0.917 & - \\
k41r & 2.761 & 10 & 0.847 & 0.865 & - \\
k46r & 2.758 & 11 & 0.939 & 0.948 & - \\
k32r & 2.758 & 12 & 0.849 & 0.873 & - \\
k72r & 2.756 & 13 & 0.915 & 0.924 & - \\
k35r & 2.747 & 14 & 0.863 & 0.888 & - \\
k84r & 2.745 & 15 & 0.861 & 0.882 & - \\
\bottomrule
\end{tabular}
}
        \caption{\textbf{Top 15 strategies in the Stewart and Plotkin tournament.}
        We report the mean score, rank, and mean cooperation rate. For comparison,
        we also include the cooperation rate from our reproduced tournament,
        alongside the cooperation rates originally reported by Stewart and Plotkin.}
        \label{tbl:sp_tournament_rankings_top_fifteen}
\end{table}

The \AXL{} library contains more than 200 strategies and regularly hosts
the largest \IPD{} tournaments to date (each time a new strategy is contributed,
the results are updated). We expand this tournament further by
including all of the original Fortran strategies (omitting the Python duplicates). 
This produces a tournament with
272 strategies.

\begin{figure}[!hbtp]
    \centering
    \includegraphics[width=.90\textwidth]{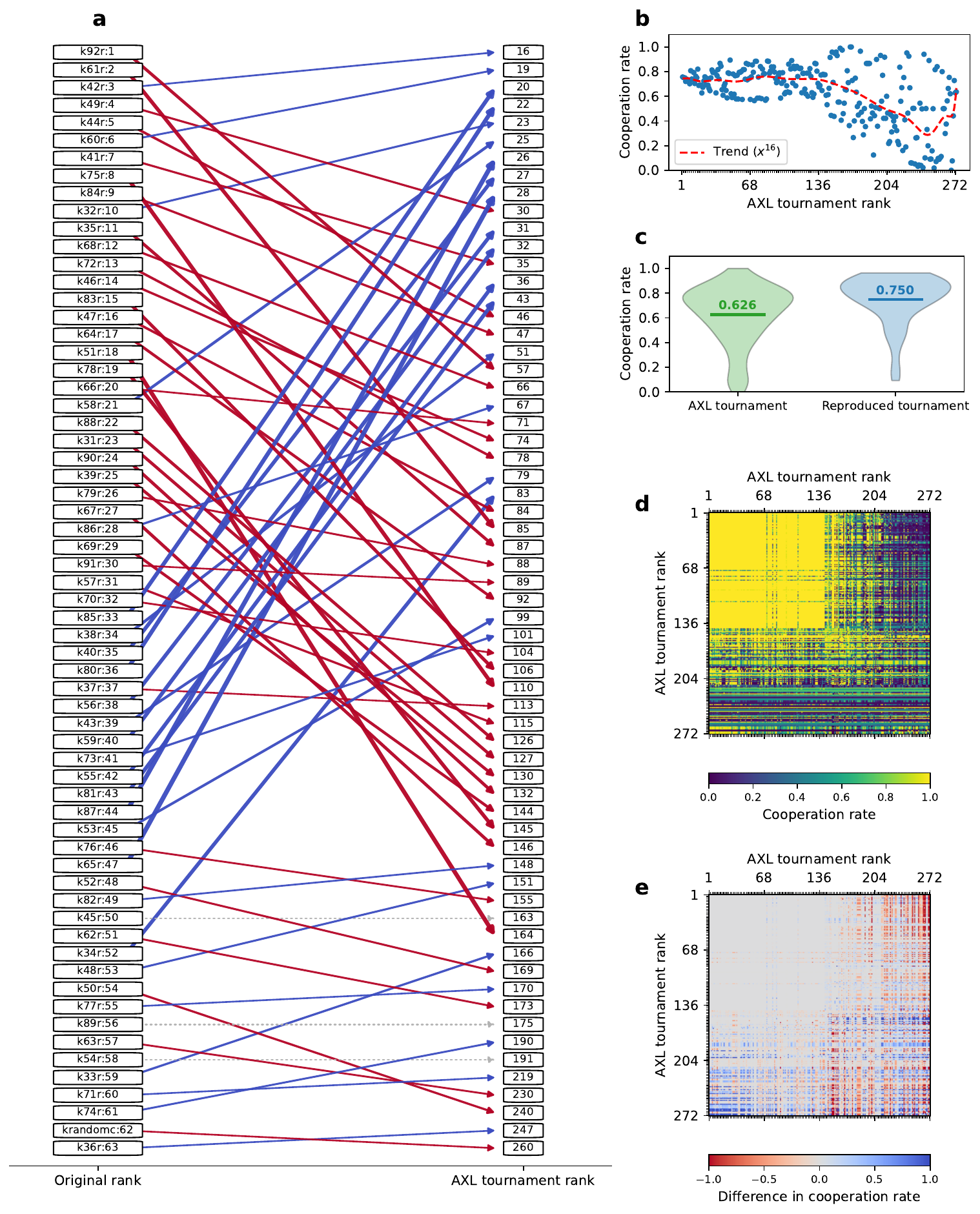}
    \caption{\textbf{Reproduction of Axelrod's second tournament with the addition
    of the strategies from the \AXL.}
    Similar to Figure~\ref{fig:rank_change_in_sp_tournament}.}
  \label{fig:rank_change_in_axelrod_python_tournament}
\end{figure}

Table~\ref{tbl:full_tournament_rankings} shows the top 16 strategies of the tournament.
Rank changes relative to the reproduced tournament are shown in
Figure~\ref{fig:rank_change_in_axelrod_python_tournament}\textbf{a}.
Several mid performing strategies from the original tournament rise into top
positions in this more complex environment. Notably, \texttt{k42r} performs
best among the original strategies, ranking 16th overall. The cooperation rate
of the library-only tournament (excluding the original Fortran strategies) is
0.619\unskip.
By contrast, the overall cooperation rate of the expanded tournament that includes
the Fortran strategies is 0.626
(see Figure~\ref{fig:rank_change_in_axelrod_python_tournament}\textbf{c}).

Pairwise interactions are shown in
Figures~\ref{fig:rank_change_in_axelrod_python_tournament}\textbf{d}--\textbf{e},
where top strategies are seen to cooperate strongly with one another, but do not
when facing lower ranking opponents. The new winners appear to exploit weaker
strategies, which contributes to their success.
The highest ranking strategies overall are sophisticated learners trained via
reinforcement learning. The ``Evolved'' strategies were trained with evolutionary
algorithms, while the ``PSO'' strategies were trained using particle swarm
optimisation, as described in~\cite{Harper2017}. These high performing
strategies do not necessarily follow Axelrod's original guidelines for success.
Instead, they align more closely with the principles identified in
\cite{glynatsi2024properties}, which are as follows:

\begin{enumerate}
    \item Be a little envious.
    \item Be ``nice'' in non-noisy environments or when games are long (if
    known).
    \item Reciprocate both cooperation and defection appropriately: be
    provocable in short matches and generous in noisy settings.
    \item It is acceptable to be clever.
    \item Adapt to the environment; adjust to the mean tournament cooperation
    level.
\end{enumerate}

\begin{table}[!hbtp]
    \centering
    \resizebox{.9\textwidth}{!}{
    \begin{tabular}{lccccccc}
\toprule
 & \makecell[c]{Average\\Score} & \makecell[c]{Rank} & \makecell[c]{Average\\Cooperation Rate} & \makecell[c]{Original Rank} & \makecell[c]{Reproduced Average\\Cooperation Rate} & \makecell[c]{Library Rank} & \makecell[c]{Library Average\\Cooperation Rate} \\
\midrule
EvolvedLookerUp2\_2\_2\textsuperscript{\textdagger} & 2.802 & 1 & 0.756 & - & - & 1 & 0.736 \\
Evolved HMM 5\textsuperscript{\textdagger} & 2.788 & 2 & 0.751 & - & - & 2 & 0.742 \\
Omega TFT: 3, 8 & 2.786 & 3 & 0.750 & - & - & 12 & 0.725 \\
Evolved ANN 5\textsuperscript{\textdagger} & 2.784 & 4 & 0.720 & - & - & 3 & 0.713 \\
Evolved ANN\textsuperscript{\textdagger} & 2.782 & 5 & 0.739 & - & - & 5 & 0.727 \\
Evolved FSM 16\textsuperscript{\textdagger} & 2.780 & 6 & 0.731 & - & - & 4 & 0.713 \\
Evolved FSM 16 Noise 05\textsuperscript{\textdagger} & 2.776 & 7 & 0.726 & - & - & 6 & 0.717 \\
PSO Gambler 2\_2\_2\textsuperscript{\textdagger} & 2.760 & 8 & 0.702 & - & - & 7 & 0.692 \\
Original Gradual & 2.757 & 9 & 0.808 & - & - & - & - \\
PSO Gambler 2\_2\_2 Noise 05\textsuperscript{\textdagger} & 2.756 & 10 & 0.740 & - & - & 14 & 0.723 \\
PSO Gambler Mem1\textsuperscript{\textdagger} & 2.756 & 11 & 0.734 & - & - & 9 & 0.728 \\
Evolved FSM 4\textsuperscript{\textdagger} & 2.752 & 12 & 0.793 & - & - & 10 & 0.777 \\
PSO Gambler 1\_1\_1\textsuperscript{\textdagger} & 2.752 & 13 & 0.704 & - & - & 8 & 0.702 \\
Gradual & 2.746 & 14 & 0.763 & - & - & 15 & 0.793 \\
DBS: 0.75, 3, 4, 3, 5 & 2.739 & 15 & 0.750 & - & - & 11 & 0.739 \\
k42r & 2.735 & 16 & 0.820 & 3 & 0.916 & - & - \\
\bottomrule
\end{tabular}
}
    \caption{\textbf{Top 16 strategies in the \AXL{} tournament.} This
    tournament includes all strategies from the \AXL{} library alongside the
    original Fortran strategies. In total, there are 272 strategies competing.
    Strategies denoted by $^\dagger$ were not designed but, they were trained
    via reinforcement learning, as described in~\cite{Harper2017}.}
    \label{tbl:full_tournament_rankings}
\end{table}

\begin{figure}[!hbtp]
    \centering
    \includegraphics[width=.90\textwidth]{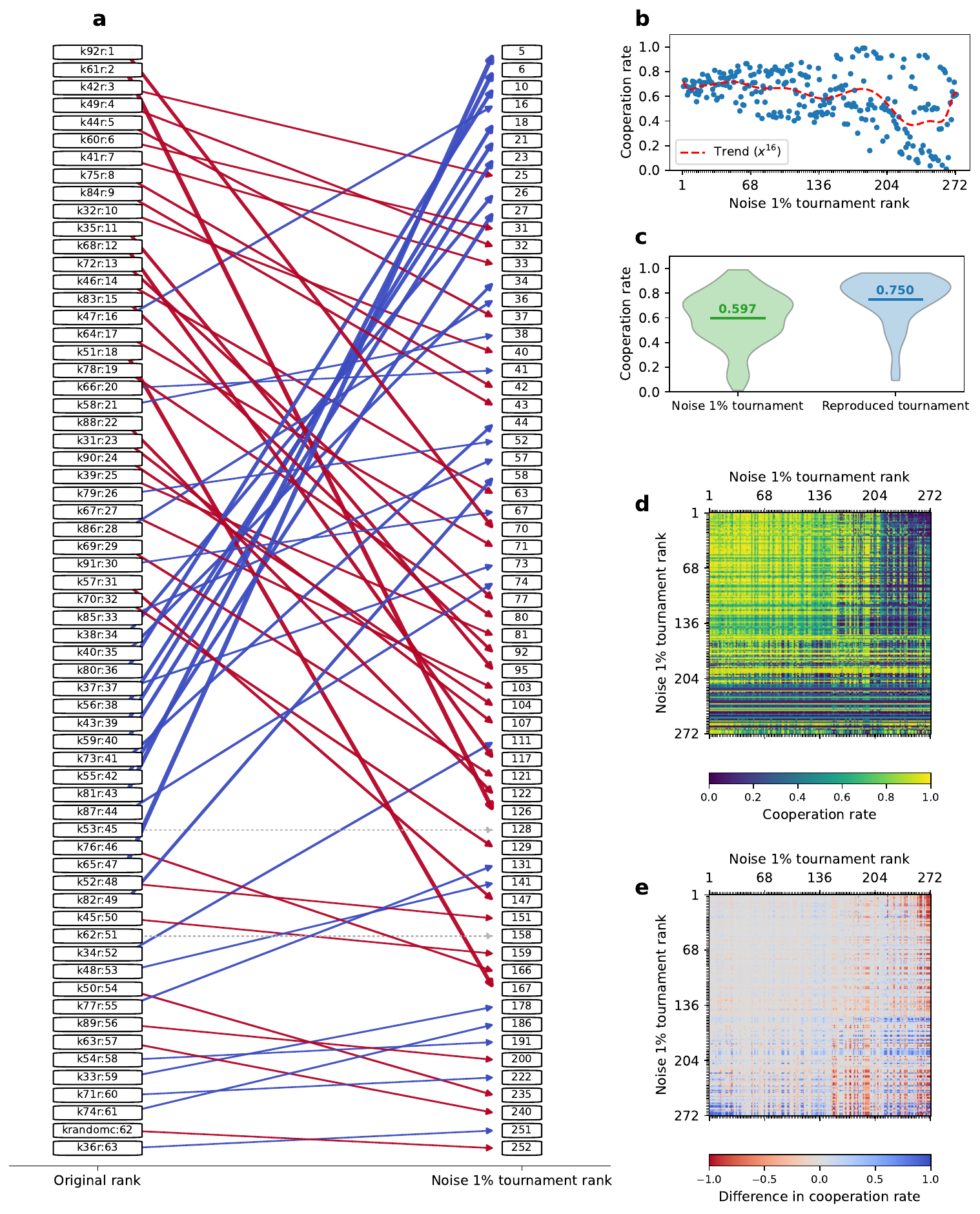}
    \caption{\textbf{Reproduction of Axelrod's second tournament with the addition
    of the strategies from the \AXL and with noise 1\%.}
    Similar to Figure~\ref{fig:rank_change_in_sp_tournament}.}
  \label{fig:rank_change_in_axelrod_python_tournament_noise_1}
\end{figure}

\begin{figure}[!hbtp]
    \centering
    \includegraphics[width=.90\textwidth]{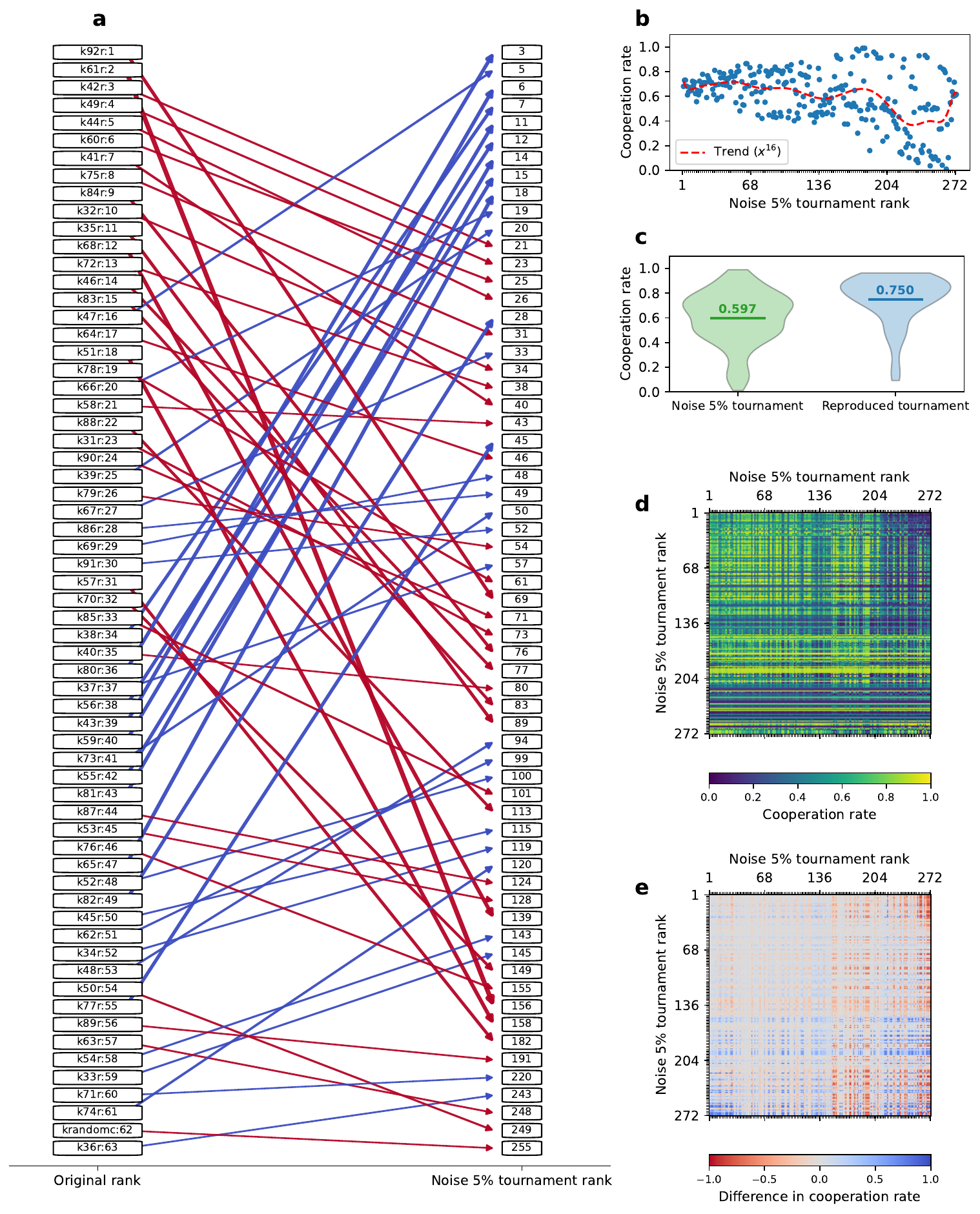}
    \caption{\textbf{Reproduction of Axelrod's second tournament with the addition
    of the strategies from the \AXL and with noise 5\%.}
    Similar to Figure~\ref{fig:rank_change_in_sp_tournament}.}
  \label{fig:rank_change_in_axelrod_python_tournament_noise_5}
\end{figure}

Finally, we consider the \AXL{} tournament with noise, using noise probabilities
of \(1\%\) and \(5\%\). The results are shown in
Figures~\ref{fig:rank_change_in_axelrod_python_tournament_noise_1} and
\ref{fig:rank_change_in_axelrod_python_tournament_noise_5}. As expected, the
overall cooperation rate is lower than in the noiseless, reproduced, and
reproduced with noise tournaments. Notably, several of the original submissions
perform very well: the best performer from the original set ranks 5th with
\(1\%\) noise and 3rd with \(5\%\) noise. This suggests that some of the more
sophisticated strategies submitted to Axelrod's original study were perhaps too
complex for the highly cooperative, relatively stable conditions where
``cooperate, retaliate after defection, forgive after cooperation, and avoid
exploiting others'' was optimal. In the presence of errors, and against
more complex competitors, these strategies make use of their additional
mechanisms to maintain competitive payoffs and robust rankings.

\section{Discussion}\label{sec:discussion}

Our study revisits one of the most influential computational experiments in the
social sciences: Axelrod's second Iterated Prisoner's Dilemma (\IPD) tournament.
By using the original Fortran implementations available online and ensuring
that they still compile, we provide the first systematic reproduction of
Axelrod's results. Despite minor discrepancies, most notably a bug we identified
in \emph{Champion} (\texttt{k61r}), our analyses confirm the robustness of Tit
For Tat (\TFT{}) as a leading strategy. This reinforces Axelrod's conclusion
that reciprocity, moderated by the capacity for retaliation and forgiveness, is
an effective principle for sustaining cooperation.

By expanding the tournament to include additional strategies from the
literature, we reveal clearer patterns. Namely, \TFT{} is remarkably robust and
difficult to displace when the majority of opponents are drawn from the original
pool. In other words, Axelrod's tournament, and similar ones, such as those
organized by Stewart and Plotkin~\cite{Stewart2012}, create highly cooperative
environments in which top performing strategies tend to reciprocate with one
another, conditions that are especially favorable to \TFT.

The picture changes in more diverse or noisy contexts. Several strategies
that ranked only moderately in Axelrod's original results rise substantially
under increased complexity or noise. A prominent example is \texttt{k42r}, which
wins 35\% of replications of the original tournament in our experiments. This
strategy outperforms \TFT{} in the presence of noise and surpasses any of the
original submissions in tournaments with more elaborate rule sets, such as the
full \AXL{} set of 209 strategies. Other mid performing strategies from the
original tournament likewise move into top positions in these settings,
particularly when noise is introduced. These findings suggest that some
sophisticated designs may have been ``over-engineered'' for the cooperative,
relatively stable conditions of the original experiment, yet prove better
adapted to noisy or diverse environments.

This observation aligns with a broader literature showing that \TFT{} is not
universally optimal. Variants such as
Win--Stay--Lose--Shift~\cite{nowak1993strategy}, Generous
\TFT{}~\cite{nowak1992tit}, and ``All-or-nothing''~\cite{hilbe2017memory}
strategies, as well as ``friendly rivals''~\cite{murase2020five} and strategies
discovered via reinforcement learning~\cite{Knight2017}, can outperform \TFT{}.
Our contribution is to show that, even within Axelrod's original submission set,
strategies that received little attention due to modest performance in the
original tournament can excel when the environment is more diverse, or noisier.
These results also help revise the common impression, shaped by the original
tournament, that \TFT{} is universally the best performing strategy.

A second contribution of this work is methodological. To our knowledge, this is
the first effort to package and reproduce, according to contemporary best
practices, code originally written in the 1980s. The archived materials~\cite{knight_2025_17250038} (at
\url{https://doi.org/10.5281/zenodo.17250038})
are curated to high standards of reproducible research~\cite{wilson2014} and
accompanied by a fully automated test suite. All changes to the original code
were made systematically and transparently, with complete records available at
(\url{https://github.com/Axelrod-Python/axelrod-fortran}).
More broadly, it serves as a case study in computational reproducibility: Axelrod's highly cited,
foundational tournaments were built on fragile digital artifacts, of which only
fragments survive. Without systematic preservation, much of this
historical work might have been lost. By updating, documenting, and integrating the
surviving code into an open-source, test-driven framework, we help ensure
continued access for the research community.

In conclusion, Axelrod's tournaments remain landmark contributions to the study
of cooperation, but their lessons are more context dependent than is often assumed.
While \TFT{} is strikingly robust in cooperative
environments resembling the original tournament, more complex strategies can
outperform it in heterogeneous or noisy settings. We hope that the field will
continue to develop preservation pipelines that bring historical computational
experiments into alignment with modern standards of reproducibility, and to
advance reproducible software and science.

\bibliographystyle{naturemag}
\bibliography{./bibliography.bib}

\end{document}